\documentclass[graybox]{svmult}

\usepackage{mathptmx}       
\usepackage{helvet}         
\usepackage{courier}        
\usepackage{type1cm}        
\usepackage{makeidx}         
\usepackage{graphicx}        
\usepackage{multicol}        
\usepackage[bottom]{footmisc}

\makeindex             

\begin{document}

\title*{The Large Magellanic Cloud: A power spectral analysis of Spitzer images}
\author{Iv\^anio Puerari, David L. Block, Bruce G. Elmegreen \& Fr\'ed\'eric Bournaud}
\institute{Iv\^anio Puerari \at INAOE, Santa Mar\'\i a Tonantzintla, Mexico; School of Computational and Applied Mathematics, University of the Witwatersrand, Johannesburg, South Africa, \email{puerari@inaoep.mx}
\and David L. Block \at School of Computational and Applied Mathematics, University of the Witwatersrand, Johannesburg, South Africa \email{david.block@wits.ac.za}
\and Bruce G. Elmegreen \at IBM T.J. Watson Research Center, New York, USA \email{bge@us.ibm.com}
\and Fr\'ed\'eric Bournaud \at CEA, Gif-sur-Yvette, France \email{frederic.bournaud@cea.fr}}

\maketitle

\abstract*{ }

\abstract{We present a power spectral analysis of Spitzer images of
  the Large Magellanic Cloud. 
The power spectra of the FIR emission show two different power laws. 
At larger scales (kpc) the slope is $\sim-1.6$, while at smaller ones 
(tens to few hundreds of parsecs) the slope is steeper, with a value
  $\sim-2.9$. 
The break occurs at a scale $\sim 100-200$ pc. We interpret this break
  as the scale 
height of the dust disk of the LMC. We perform high resolution
  simulations with and without stellar feedback. 
Our AMR hydrodynamic simulations of model galaxies using the LMC mass
  and rotation curve, 
confirm that they have similar two-component power-laws for projected
  density $-$ 
and that the break does indeed occur at the disk thickness. 
Power spectral analysis of velocities betrays a single power law for
  in-plane components. 
The vertical component of the velocity shows a flat behavior for large
  structures and a power law 
similar to the in-plane velocities at small scales. The motions are
  highly anisotropic at large scales, 
with in-plane velocities being much more important than vertical
  ones. 
In contrast, at small scales, the motions become more isotropic.}

\section{Introduction}
\label{sec:1}

The LMC belongs to the de Vaucouleurs classification bin SB(s)m and is
$\sim 50$ kpc away (based on a 
distance modulus $m-M=18.50\pm0.10$, following Freedman et
al. 2001). The LMC is thus an ideal laboratory 
for studies of both morphology and of turbulence, due to the high spatial
resolution possible due to its close proximity. Furthermore, the LMC has
relatively little shear, 
and has no strong spiral arms:  the structure of the interstellar
medium (ISM) is less affected by density wave shocks. 

The study of the morphology of galaxies facilitates our understanding
of the physical processes of formation and of secular evolution 
which forge the distribution of stars, gas, dust, and the like. 

Particularly interesting is the study of the spatial structure of the ISM.
Atomic hydrogen HI gas is a sensitive tracer of the ISM.
Due to the fact that radio observations of nearby galaxies can attain
a good spatial
and velocity resolution, HI gas distributions may been analysed
in terms of power spectral analysis (Crovisier
\& Dickey 1983, Green 1993, Stanimirovic et al. 1999,
Elmegreen et al. 2001, Dutta et al. 2008, 2009a,b).

Another tracer of the structure of the ISM is cold dust, in the form
of carbonaceous and silicate grains, at temperatures 
of $10-20$ K. These grains are the dominant emitters in the FIR.
Li \& Draine (2001) have extensively modeled COBE/FIR observations of
the diffuse ISM in the Milky Way. Li \& Draine (2002) focus 
their attention on the SMC and show that carbonaceous and silicate 
grains of different sizes
dominate the spectra at wavelengths in the 
range 50 to \hbox{400 $\mu$m.}
Large ($a>250$ \AA) grains become more important for wavelengths
larger than 50 \AA, while small carbonaceous grains prevails at shorter
wavelengths, where one observes PAH's lines.

In this study, we present a power spectral study conducted on images
of the LMC secured with the Spitzer Space Telescope and MIPS.
We analyse images at  24, 70 and \hbox{160 $\mu$m}.
The longest of these wavelengths 
traces the dominant component of interstellar
matter: cold dust grains. 
At 70 $\mu$m the emission arises from both
warm and cold dust grains, while at 24 $\mu$m the emission is from
warm dust grains and PAH's.
Cold dust grains are responsible for the extinction in optical images,
the scattering of starlight, as well as polarization. 
It is reasonable to state that
in studying emission longward of 60 $\mu$m, one is investigating the
structure of the optical mask itself.

\section{Data}
\label{sec:2}

The
coverage in the MIPS images of the LMC is
$\sim 8 \times 8$ degrees, with a total integration time of 217
hours. The point source sensitivity estimates for the SAGE survey improve previous
surveys of the LMC (such as those with the Infrared Astronomy Satellite
IRAS) by three orders of magnitude (Meixner et al. 2006).

At 24 $\mu$m and 70 $\mu$m, the images sizes are  8192 $\times$ 8192
pixels, at a scale of 4.98$''$ and 4.8$''$ per pixel, respectively.
At 160 $\mu$m, the image size is 2048 $\times 2048$ pixels, at a scale of
15.6$''$ per pixel. As far as deprojecting images of the LMC are
concerned,  there is no unique ``center'' - the dynamical center of the
HI is offset by almost a full degree from the photometric center of the
LMC bar (Westerlund 1997).  We choose to  conduct our deprojections
about the dynamical centre of the LMC, which has right ascension
$\alpha = 5^{h} 27.6^{m}$ and declination $\delta = -69^{\circ}52^{'}$
(J2000.0), following section 7 in van der Marel et al. (2002). In order
to test the robustness of our analysis, we also deprojected the MIPS
images about the bar center as opposed to the dynamical center of the
LMC; the resulting power spectra are only marginally affected.

\section{Analysis}
\label{sec:3}

A power spectral calculation is straightforward. In two dimensions,
the prescription is as follows;

\begin{equation}
{\cal F}(k_x,k_y)=\int_x \int_y I(x,y)e^{-i(k_xx + k_yy)}dxdy
\end{equation}

\noindent where $I(x,y)$ is the intensity of each pixel, and $k_x$ and $k_y$ are the
conjugate variables to $x$ and $y$, respectively. The full 2D power
spectra is given by

\begin{equation}
Power(k_x,k_y) = (Re {\cal [F]})^2 + (Im {\cal [F]})^2
\end{equation}

A 1D PS $Power(k)$ can be calculated using $k^2=k_x^2+k_y^2$. Mathematically,
this corresponds to take circles in the Fourier space ${\cal F}$.

\section{Results}
\label{sec:4}

In Figure 1, the left hand panel shows emission from LMC dust at 160
 $\mu$m.
At top right is seen 
the Fourier space ${\cal F}$ while the power spectrum itself 
$Power(k)$ is shown at the lower right of Figure 1.

\begin{figure}[ht]
\begin{center}
\includegraphics[scale=.36]{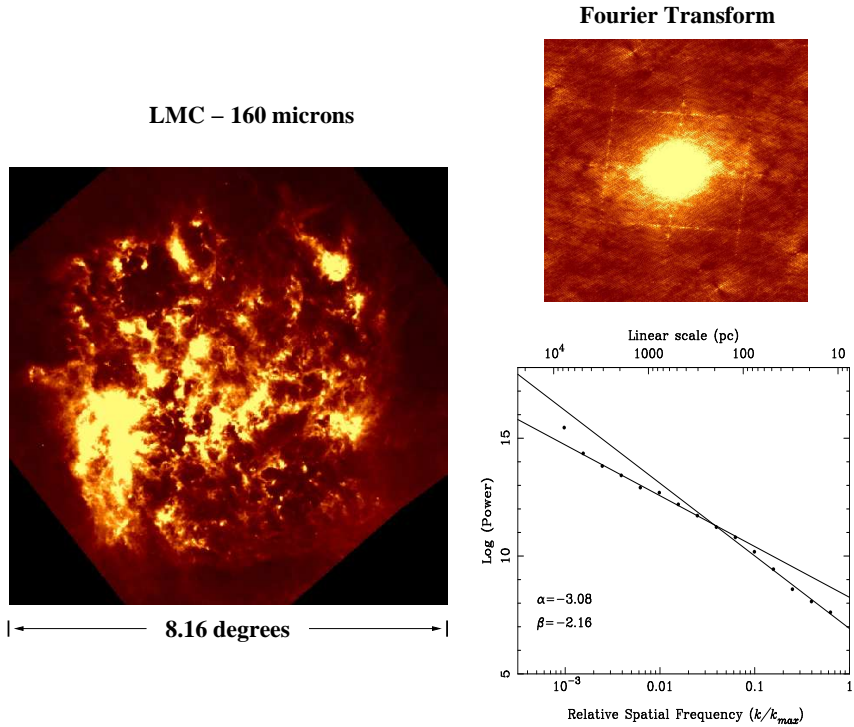}
\caption{Left Panel: The Large Magellanic Cloud imaged at 160 $\mu$m,
  as part of the SAGE survey. 
North is up, and East to the left. 
Emission from cold dust grains is seen in striking contrast. 
The bright emission cloud in the lower left of the MIPS image
corresponds to the giant molecular and 
atomic clouds south of the 30 Doradus star-forming region. 
Upper Right Panel: The two-dimensional Fourier transform of the
deprojected (``face-on'') MIPS image at 
160 $\mu$m. The rectangle betrays the presence of low level striping
in the SAGE MIPS image. 
Lower Right Panel: Two-dimensional power spectrum of the Large
Magellanic Cloud.  
The power spectrum was divided into fifteen intervals of relative
spatial frequency, 
and each dot represents the mean of the power spectrum in that
interval. 
Least-squares regression lines are used to determine the slopes of the
two distinct power laws. 
The power-laws are $P(k) \propto k^\alpha$ and $P(k) \propto k^\beta$
for high and low spatial frequencies, 
respectively. The point of intersection of the two regression lines at
$\sim$ 200 pc demarcates the ``knee'', 
interpreted as the line-of-sight depth of the cold dust disk of the Large Magellanic Cloud.}
\label{fig:1}
\end{center}
\end{figure}

\begin{figure}[ht]
\begin{center}
\includegraphics[scale=.36]{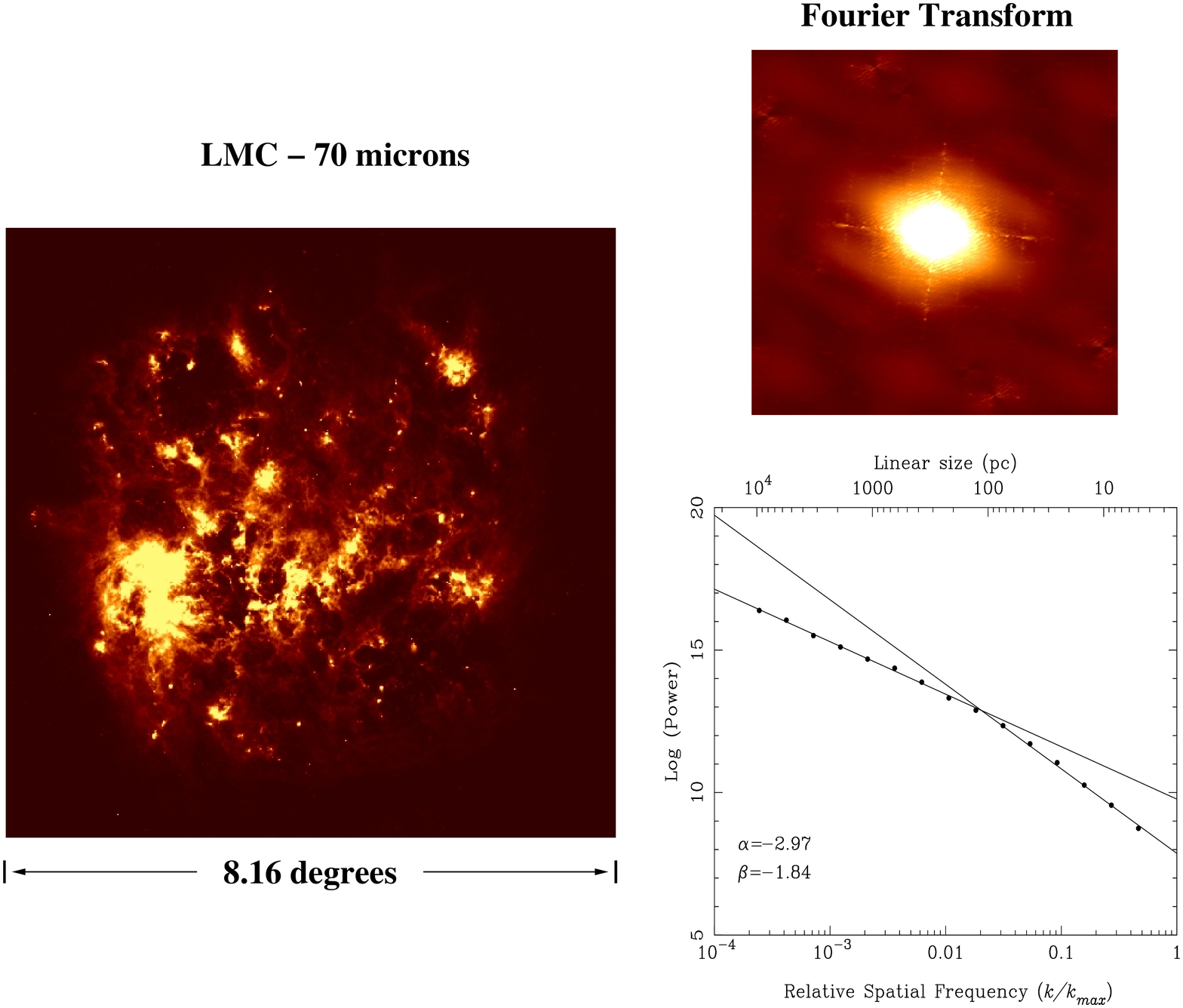}
\caption{Two-dimensional power spectral analysis of the Large
  Magellanic Cloud at 70 $\mu$m. 
The original 70 $\mu$m image appears in the left panel. 
Details as in Figure 1. Shown in the lower right hand panel are two
power laws, 
and an associated ``knee'' which occurs between 100 and 200 pc. 
Particularly striking is the smallest scale sampled in the 70 $\mu$m
power spectrum of only 2.32 pc, 
at the Nyquist limit. The power spectrum spans almost 4 orders of
magnitude. 
The value of the steeper (more negative) slope of $-2.97$ in the power
spectrum is close to the $-3$ slope 
representative of 2D incompressible Kolmogorov turbulence.}
\label{fig:2}
\end{center}
\end{figure}

The 2D power spectrum of the LMC at 160 $\mu$m demonstrates two
distinct power laws (see Block et al. 2010 for further details).
The steeper least-squares slope at higher spatial
frequencies has a value of $-3.08$, while the shallower least-squares
slope is $-2.16$.   A distinct `knee' or `break' in the
power spectrum is seen at approximately 200 parsecs. Figure 2 shows a
very similar behaviour -- this time for the 70$\mu m$ image. 
At all three MIPS wavelengths, the knee is found to lie in the
interval spanning the $100-200$ pc. The least-square slopes 
at the three different MIPS wavelengths are similar, but with
a slight progression, as follows: at high spatial frequency, the
values are $-3.08$,
$-2.97$ and $-2.60$ at 160 $\mu$m, 70 $\mu$m and 24 $\mu$m,
respectively, while at low spatial frequency, they are computed to be
$-2.16$, $-1.84$
and $-0.80$ in these three passbands. 
The implication may mean that cool dust
is more diffuse or dispersed than hot dust.

The spatial resolution of the MIPS images allows one to probe
structures of dust emission at both small and large physical scales. 
At small scales (smaller than the disk scale height), the behavior of the
power spectra is a 3D one, while for large scales, the PS is more
akin to that in 2D. The LMC disk is relatively thin, 
with a height/size ratio of $\sim 1/15-1/20$; nevertheless, motions do
not occur in strictly 2D, as in an infinitely thin 2D sheet. 

We have no information on the actual velocities seen in the
LMC images, but we believe that
two different physical process are acting in the distribution of the ISM
at large and small scales.  Conceivably, more localized energy sources (from stars, OB associations
and SN explosions) should drive 3D motions and contribute to the small scale sector
of the power spectrum.  In contrast, density-wave and bar-driven streaming motions are often
5 to 10 times faster than the perpendicular motions which produce the disk thickness
and which are thus responsible for the 2D behavior of the ISM distribution.

\begin{figure}[ht]
\begin{center}
\includegraphics[scale=.45]{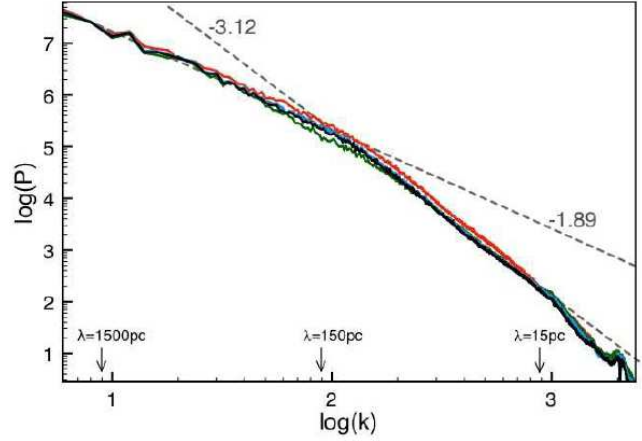}
\caption{Density power spectrum for 4 different times in the simulation with feedback. T = 254 Myr (dark), 261 Myr (green), 268 Myr (blue) and 275 Myr (red).}
\label{fig:4}
\end{center}
\end{figure}

\begin{figure}[ht]
\begin{center}
\includegraphics[scale=.45]{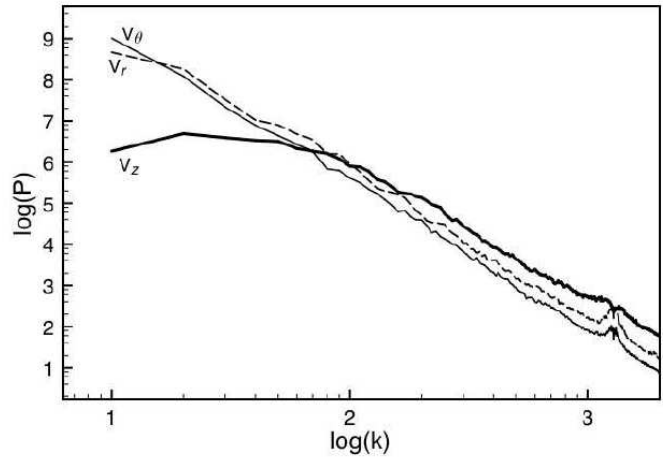}
\caption{Velocity power spectrum for the three separate components $V_R$, $V_\theta$ and $V_Z$, at T = 254 Myr (LMC model with stellar feedback).}
\label{fig:5}
\end{center}
\end{figure}

\section{Simulations}
\label{sec:5}

We have performed high spatial and mass resolution hydrodynamic
simulations
(up to 0.8 pc and
$5\times10^3$ M$_\odot$), to study the
properties of the ISM substructure and turbulence of LMC models.
We use an AMR (Adaptive Mesh Refinement) software written by
Romain Teyssier (see Teyssier 2002).
A complete discussion appears in Bournaud et al. (2010);
in this present paper, we highlight some details. We initialize an exponential
stellar disk containing $3\times10^9$ M$_\odot$, a scalelength of 1.5 kpc
and a truncation radius of 3 kpc. A non-rotating bulge ($3\times10^8$ M$_\odot$) and
a halo ($5\times10^9$ M$_\odot$) are both included.  The initial gas disk is
exponential, with a scalelength of 3.0 kpc, and a scaleheight of 50 pc (truncated
at 500 pc). The total gas mass is $6\times10^8$ M$_\odot$. Initially, the gas is purely
rotating, with no macroscopic velocity dispersions.  With time, gravity generates
turbulence at several scales. For purposes of comparison, we generated
two models:  with and without the feedback from star formation.

In Fig. 3 we show the power spectra of the gas distribution for the `with 
star formation feedback' run at times T=254 Myr and at three subsequent times,
incremented by 7 Myr. The two slopes are clearly seen, and their values are $-3.12$ for
large scales, and $-1.89$ for small scales. These values agree with
those calculated observationally, using the MIPS FIR images above. 
We have also computed power spectra of the velocities 
(in-plane $V_R$ and $V_\theta$, and vertical $V_Z$)
(seen in  Fig. 4).  At small scales, all of the PS of velocities have
the same slope with a marginal difference in power.  At large
scales, the motions are strongly non-isotropic, with the $V_Z$ power spectra being
flat, with less power, compared to in-plane velocities. The latter ones present the
same slope at high and low frequencies. This result is significant, because it
demonstrates that the vertical component of the velocity 
behaves very different for high and
low spatial frequencies. 
At high frequencies, the turbulence is more 3D (almost the same
power for $V_R$, $V_\theta$, and $V_Z$). At low frequencies (large
scales), however, the
motions are 2D (much less power for $V_Z$ than for the in-plane
velocities). 
These
2D streams, generated primarily by disk self-gravity --  but possibly also by tidal
forces from companion galaxies, as in Mastropietro et al. (1999) and
from intergalactic ram 
pressure, as discussed by Tonnesen \& Bryan (2009) and by Dutta et
al. (2010) -- contribute to the low
frequency portion of the power spectrum.


\section{Conclusions}
\label{sec:6}

For the first time, power spectra are presented for emission for dust
grains  in the LMC. This is facilitated by analysing 
FIR images at wavelengths of 24, 70, and \hbox{160 $\mu$m} from MIPS, on board the Spitzer
Space Telescope. The power spectra reveal two distinct power laws in
the high and
low frequency domain, representing small and large physical scales,
respectively. 
The `knee' or `break' in the power
spectra occurs at $\sim 100-200$ pc. We interpret this break as the scale height of the
dust disk of our closest magellanic neighbor, the LMC,
following Elmegreen, Kim, \& Staveley-Smith (2001). 
High resolution AMR simulations of LMC models affirm that the
same behavior 
follows in the power spectra generated from our simulated ISM:
simulations confirm the existence of two different power
laws, with the `break' occurring at the scale height of the
disk. Observations and simulations are in excellent accord. 
As far as velocity components are concerned,
we find that for high frequencies (small scales), the motions are quite isotropic, while
at low frequencies, in-plane motions are much more important than vertical ones.  These 
results affirm that the power spectra of the ISM does indeed depend
upon the geometry of the system being simulated.
At large scales, perturbations create flows akin to those in two dimensions, 
while at small scales, simulations betray a behavior which is
more three dimensional in nature. 

\begin{acknowledgement}
IP extends his thanks to the Mexican Foundation Conacyt. 
He also expresses profound gratitude to the School of
Computational and Applied Mathematics at the University of the
Witwatersrand, Johannesburg, for their hospitality during his many visits
to the University, spanning more than twelve years.
\end{acknowledgement}
 
{}

\end{document}